\documentclass[11pt]{article}
\usepackage{amsmath,amssymb}
\usepackage{hyperref}

\newsavebox{\uuunit}
\sbox{\uuunit}
    {\setlength{\unitlength}{0.825em}
     \begin{picture}(0.6,0.7)
        \thinlines
        \put(0,0){\line(1,0){0.5}}
        \put(0.15,0){\line(0,1){0.7}}
        \put(0.35,0){\line(0,1){0.8}}
       \multiput(0.3,0.8)(-0.04,-0.02){10}{\rule{0.5pt}{0.5pt}}
     \end {picture}}

\usepackage[table]{xcolor}

\usepackage{graphicx}
\usepackage{bbm}

\newcommand{\ba}{\begin{eqnarray*}}
\newcommand{\ea}{\end{eqnarray*}}
\newcommand{\ban}{\begin{eqnarray}}
\newcommand{\ean}{\end{eqnarray}}

\def\beq{\begin{equation}}
\def\bee{\begin{equation}}
\def\eeq{\end{equation}}
\def\bea{\begin{eqnarray}}
\def\eea{\end{eqnarray}}
\def\bd{\begin{displaymath}}
\def\ed{\end{displaymath}}

\numberwithin{equation}{section}

\begin{document}

\thispagestyle{empty}
{}


\vskip -3mm
\begin{center}
{\bf\LARGE
\vskip - 1cm
 Band Structure in Yang-Mills Theories\\ [2mm]}

\vspace{10mm}

{\large
{\bf Constantin Bachas$^{1}$ and Theodore Tomaras$^{2}$ }
  \vskip 3mm

\vspace{1cm}

{\it 
$^1$ Laboratoire de Physique Th\'eorique de l' \'Ecole Normale Sup\'erieure,\\ 
 PSL Research University,  CNRS,  Sorbonne Universit\'es, UPMC Univ.\,Paris 06,\\ 
24 rue Lhomond, 75231 Paris Cedex 05, France  \\ [4mm]
 $^2$ Department of Physics and Institute for Theoretical and Computational Physics, University of Crete, 70013 Heraklion, Greece
}}

\vspace{5mm}

\end{center}
\vspace{5mm}

\begin{center}
{\bf ABSTRACT}\\
\end{center}
 
    We show how Yang-Mills theory on S$^3\times\mathbb{R}$
 can exhibit a   spectrum with continuous bands if 
 coupled either  to a topological 3-form gauge field,   or   to
a dynamical axion with heavy Peccei-Quinn scale. 
 The basic mechanism consists in associating  winding histories to  a bosonic zero mode whose role is to
convert  a circle in configuration space into a helix. The  zero mode is,  respectively,
 the   holonomy of the 3-form field or the axion momentum. 
In these models different $\theta$ sectors coexist but  are not mixed
by local operators. 
Our analysis sheds light on, and extends  Seiberg's proposal for modifying the topological sums
in quantum field theories. It  refutes a recent claim that $B+L$ violation
at LHC is unsuppressed.

\vskip 7.7cm

\rightline{LPTENS 16/01}
\rightline{ITCP-IPP 2016/03}

\clearpage
\setcounter{page}{1}

\section{Introduction}

   While revisiting recently the question of electroweak $B+L$ violation  \cite{'tHooft:1976up}    in collider experiments
(for reviews and more references see  \cite{Bezrukov:2003qm}\cite{Ringwald:2003ns}\cite{Mattis:1991bj}\cite{Guida:1993qy})
Tye and Wong  made 
a bold proposal  \cite{Tye:2015tva}\cite{Tye:2016pxi}.  They  argued that
the relevant degree of freedom in the space-time interaction region for such processes
 is the Chern-Simons number of the electroweak gauge fields, whose effective quantum mechanics
exhibits band structure similar to that of an electron in a solid. They further argued that the
width of these bands is proportional to the amplitude of the instanton-induced tunneling transitions which
grows with available energy, and  went on to conclude that $B+L$ violation may be 
unsuppressed at LHC. 

       This last conclusion 
 misses in our opinion the key aspect of the problem:  the difficulty of streamlining
the collision energy into  coherent excitations  of one or   few quantum degrees of freedom.
In weakly-coupled theories  such streamlining  is expected to be exponentially small, 
independently of any other features such as the existence of many vacua. 
Thus, even though the inclusive cross-section  for tunneling processes does grow with collision energy \cite{Ringwald:1989ee,Espinosa:1989qn} 
  the  growth  
 is believed to stop much before one hits the unitarity bound. This can be  shown rigorously
in an analog $\lambda \phi^4$ quantum mechanical model \cite{Bachas:1991cb,Bachas:1992dw}.
\smallskip

   Quite independently  of LHC physics, the claim that Yang-Mills theory can exhibit   band structure is
by itself surprising. Energy bands arise when a particle 
 moves in a non-compact dimension  with  periodic potential  $V(q)= V(q+n), \,\, n\in \mathbb{Z}$.
A band is spanned by the quasi-momentum (or lattice momentum) which takes values
in a Brillouin zone. 
In the Yang-Mills case the coordinate $q$ is   the Chern-Simons number of the gauge field,
and the shifts  $q\to q+n$ are implemented by large gauge transformations. 
Since these are {\it gauge} and  not  {\it global} symmetries,    $q$  is identified with $q+1$ 
and the wavefunction  must  be a periodic function  of $q$. This fixes the  quasi-momentum,   so  there are no bands. 

\smallskip
   The purpose of this short  letter  will be  to discuss how to evade this conclusion. 
The basic  idea is to associate   tunneling transitions to a  bosonic zero mode which  provides
a `tag'  that distinguishes  the winding vacua. This device effectively converts large gauge transformations into
global symmetries or, more figuratively, turns the $q$-circle into a (non-compact)  helix for which the
spectrum exhibits  (continuous) bands. The  question then is whether this strategy can be
implemented within  {\it local}  quantum field theory.  We will discuss  two different ways of doing 
 this on a compact space manifold.

The first,  proposed by Seiberg \cite{Seiberg:2010qd}, 
involves coupling  Yang-Mills   theory to a topological 
 3-form gauge  field $B$. What tags the  winding vacua is
the holonomy  $b=\int_{M_3}\hskip -1.0mm B\, $ of $B$  around  the  space  manifold $M_3$. 
Another, less contrived but only approximate  way  involves
 coupling  Yang-Mills   to a dynamical
axion, $a$, with very large  decay constant $f_a$. 
The linear coordinate of the  helix is in this case 
  the axion momentum or, equivalently, the integrated  Poincar\'e dual 3-form\,\,  
$\int_{M_3}\hskip -1.4mm \,^*da$\,. As we will see, the axion  model 
(which arises naturally from  string theory) 
reduces to the topological model  
in the limit   $f_a\to \infty$.
 
{   The bands in the above  extensions of gauge theory are  filled by coexisting
 $\theta$ sectors, which  are not  mixed by local operators  but do talk via  charged-membrane 
 interface probes (3-dimensional analogs of Wilson lines). 
Even under the optimistic assumptions  (1)  that we may replace the 3-sphere in our analysis
by an interaction
region of  size $m_W^{-1}$,  and (2) that the probe
 membranes become dynamical, these latter  would be 
too heavy to play any role in  collider physics. Thus, independently of  any other objections,
the band structure that we exhibit here is not relevant for $B+L$ violation at LHC.
Nevertheless our results provide  new insights
on the topological sector  of non-abelian gauge theories, and the `helix mechanism' 
is so simple and general that it may find  applications in  other contexts.
}
\smallskip

    The  paper is organized as follows.  In section \ref{2}\, we describe the basic idea  with 
 the example of a charged particle moving   on a  helix   or on a circle, 
 in a gravitational potential. Because $2\pi$ rotations   are  global (respectively gauge)  symmetries, 
the particle's  spectrum does (does not)  have   energy bands. 
In section \,\ref{3}\, we explain  how Seiberg's proposal
to couple Yang-Mills  to a topological 3-form gauge field
  implements the helix  mechanism in local quantum field theory.  
Section \,\ref{4}\, presents a  different realization which 
involves   a dynamical axion  with high Peccei-Quinn  scale $f_a$. 
We show that  in this theory approximate $\theta$ sectors coexist, and  that 
in the limit   $f_a\to\infty$ one recovers the non-compact version of Seiberg's model.
 Finally, section \ref{5}\, contains some concluding remarks, in particular on electroweak
$B+L$ violation -- the issue that  prompted this investigation. 

A quantum-mechanical action for the Chern-Simons 
variable,  similar to the one proposed by  Tye and Wong \cite{Tye:2015tva}, 
can be derived by putting the SU(2) theory on the 3-sphere and then projecting onto the SO(4)-invariant sector.
This is an amusing exercise,   but  since the  details of this action  play no   role 
  we relegate it  to the appendix.

 
\section{Circle versus  Helix}
\label{2}

     One reads in many textbooks that Yang-Mills theory  has  winding vacua  $\vert n\rangle$,
  that the true vacuum is a linear superposition  $\vert \theta\rangle = \sum_n e^{in\theta} \vert n\rangle$, and that different  $\vert \theta\rangle$  belong to distinct superselection sectors.  
Though usually a matter of semantics, these statements  hide a subtle aspect of the problem
and can  be misleading. We start by explaining   why in 
  a simple  model of quantum mechanics. This  is standard material (see e.g.  \cite{VR}) and
 readers may want to go quickly to the next section. 
  
The model is that of  a particle moving in a periodic potential  $V(q)= V(q+n), \, n\in \mathbb{Z}$
with action
\bea\label{L}
S = \int dt\,  {\cal L}\  = \ \int dt\, \bigl[ {M\over 2} \dot q^{\,2}  - V(q)   - {\theta } \, \dot q\, \bigr]\ .  
\eea
 We have added to  ${\cal L}$ a topological  term that plays the   role of  the  $\theta$  term in
 Yang-Mills theory, as will be  clear in the following section. 
To fully specify  the model   we must still state   whether the  symmetry under 
discrete translations,  $q\to q+  n$, 
 is {\it global}
  or {\it gauged}.\,\footnote{More generally it is possible to only gauge  a subgroup $q \to q+   N n$ 
of the symmetry group  which amounts to  putting   the particle on a circle of radius $N/2\pi$.} 
 In the former case the particle lives on the real line and the potential   has an infinite number of
local minima  labelled by $\, n\in \mathbb{Z}$, while  in the latter it  lives on a circle  and
$V$ has a unique minimum which we may choose to be at $q=0$.

\begin{figure}[t!]
\centering 
\includegraphics[width=.90\textwidth,trim=0cm 6cm 0cm 6cm,clip=true]{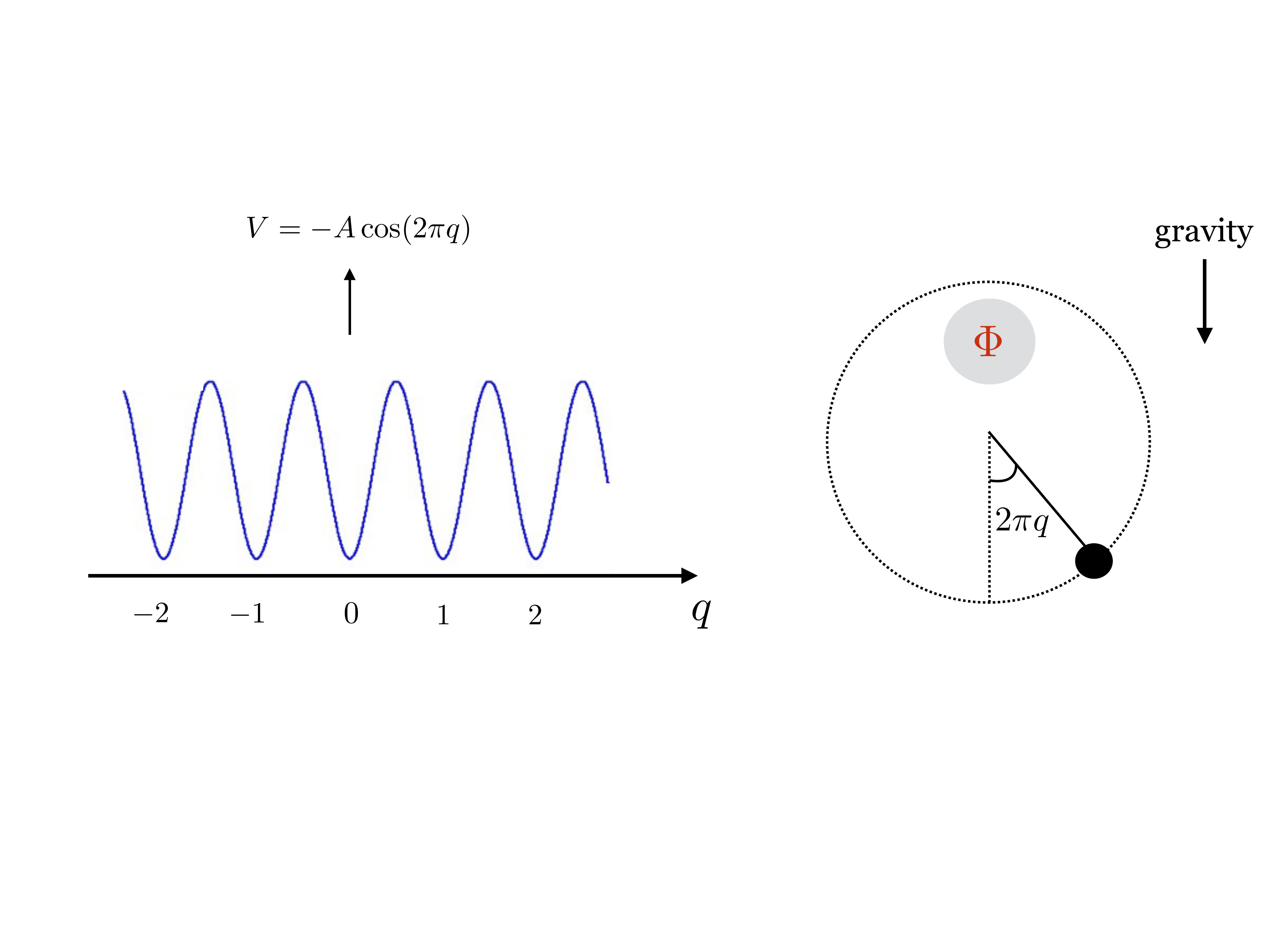}
\vskip -6mm
 \caption{\small A particle moving on the real line in the  potential $V = -A \cos (2\pi q)\,$  has the same
action as a  pendulum of unit radius  with $A = M\times$(Earth's gravity). In the pendulum the 
discrete symmetry under $2\pi$ rotations 
is gauged and $V$ has a unique minimum,  whereas  on the real line the
symmetry is global and the minima are distinct.
The  magnetic flux $\Phi$  shown in the figure  induces a theta term with
$\theta = e \Phi$,  where $e$ is  the particle's charge.  }
\label{fig:2} 
 \end{figure}

A physical realization of the circle model is a quantum pendulum moving in the Earth's
gravitational field. The topological $\theta$  term arises if  the pendulum carries  electric charge 
and  encircles a magnetic flux. To  turn  the gauge 
symmetry of  $2\pi$ rotations  into a global symmetry one can employ a simple trick:
Convert  the circle into a helix by forcing  the particle to move  in an extra (horizontal)  dimension $z$,
 so that
$z = \xi  q$\,.  The  minima of the potential  are now  distinguished  by
their position in the extra dimension,  and $q$ is effectively decompactified. The $\theta$  term  
still   corresponds to a flux tube threading  the helix.  These facts are illustrated in the figure \ref{fig:2}
above.  

  The physics of these problems is familiar. In the non-compact case  the energy eigenstates
are organized in  continuous bands,  while in the compact case one keeps  a single 
Bloch wave in every band and the spectrum is discrete. At the risk of being pedantic,  let us 
recall how this works  in more detail. 
 From    \eqref{L} one  finds  the momentum and  Hamiltonian 
\bea 
 p \equiv {\partial {\cal L}\over \partial \dot q} =  M\dot q - \theta\ , \qquad {\cal H}(\theta) \equiv  p \dot q - {\cal L}
 =\  {1\over 2M}  (p + \theta)^2  +  V(q)\ . 
\eea
The  $\theta$-dependence looks, at first sight,    trivial since  it
is  removed by  a  unitary transformation 
\bea\label{equiv}
 p +  \theta = e^{-i \theta q} \,   p\, e^{i \theta  q}\quad \Longrightarrow\quad 
 {\cal H}(\theta) =  e^{-i \theta q}\,   {\cal H}(0) \, e^{i \theta  q}\ . 
\eea
Indeed,  $e^{i \theta  q}$ is   the Bohm-Aharonov   phase for a
 particle moving around  a magnetic flux tube  as in figure \ref{fig:2}. 
However, while $e^{i \theta  q}$ is a legitimate operator in the 
 helix  model, it is 
  {\it not}   for  the pendulum  because it  is multi-valued on the circle 
(unless $\theta\in 2\pi \mathbb{Z}$). Thus $\theta$ is a relevant parameter   in the compact case,
whereas it can be absorbed by  a redefinition of momentum in the non-compact case.

\begin{figure}[htbp!]
\centering 
 \includegraphics[width=.88\textwidth,trim=0cm 5cm 0cm 5cm,clip=true]{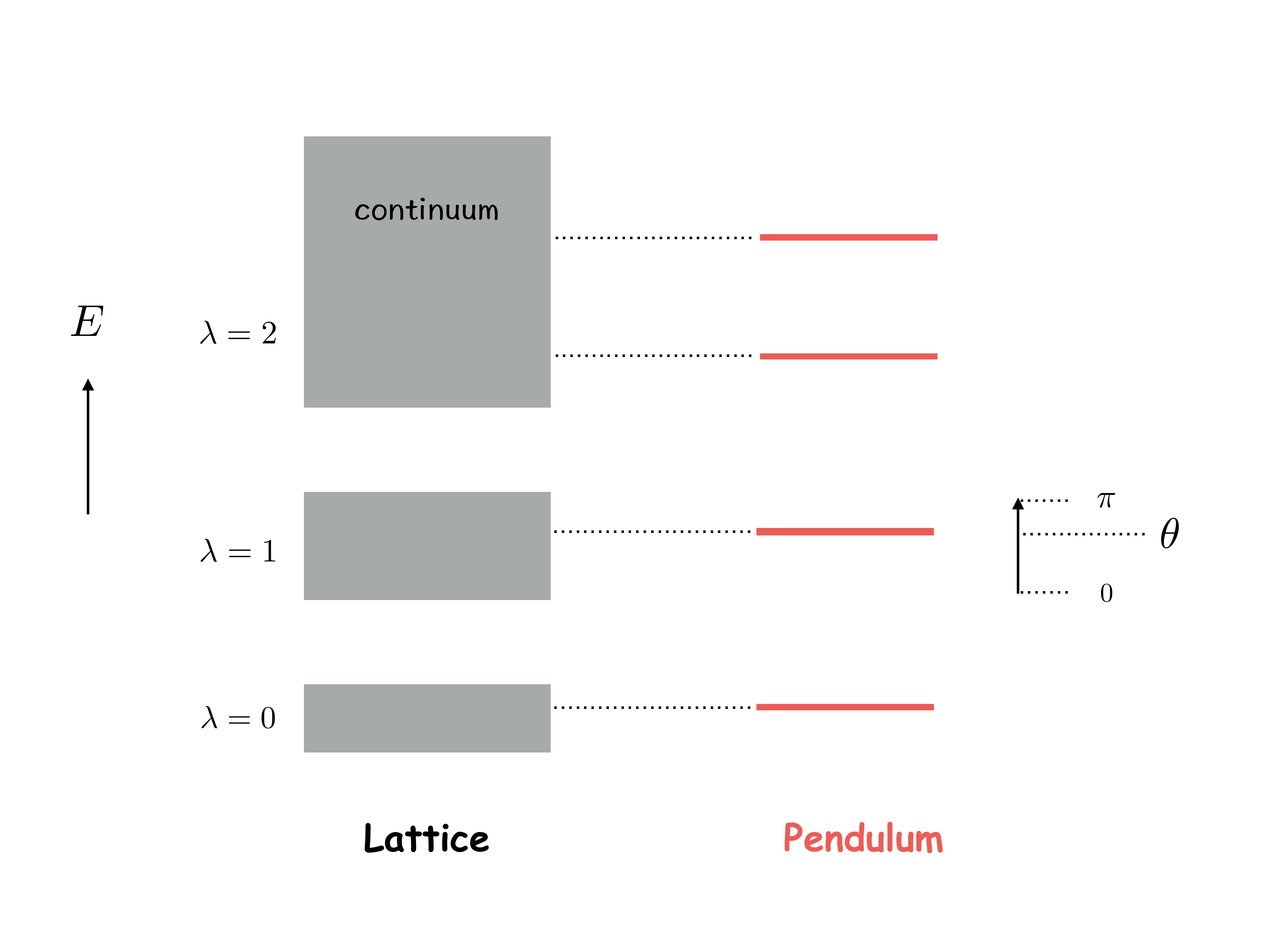}
\caption{\small  On the left the typical  band structure for a particle moving in a   periodic potential on the real line
(assuming $k\to -k$ invariance). 
The width of the bands grows with the tunneling amplitude between neighbouring   wells
and bands eventually merge in a continuum.
Compactifying   the line to a  circle projects onto  states with effective quasi-momentum $\theta$
(the red spectrum  on  the right).  
 }
\label{fig:1} 
  \end{figure}

Either way,   the generator 
 of discrete translations,  $P := e^{  i p}$, commutes
with  
 ${\cal H}$ and   can  be  diagonalized. 
Its eigenvalues are phases  $e^{  ik}$,
where $k\in [-\pi, \pi]$\,\  is  the  lattice-momentum or quasi-momentum.
Energy eigenstates of  the helix model are   labelled by   $k$ 
 and by the  index $\lambda =0, 1 , \cdots$  that designates an `electron' band. 
 A generic spectrum is  illustrated in figure \ref{fig:1}\,. 
It has   narrow `tight-binding'  bands 
whose width grows with $\lambda$,   until   they eventually  merge in   a continuum. 
If   $E_{k, \lambda}(\theta)$ and $\psi_{k, \lambda}(q, \theta)$ are the eigenvalues 
and 
 eigenfunctions  of ${\cal H}(\theta)$,   the 
 unitary equivalence \eqref{equiv} implies  
 \bea
\psi_{k, \lambda}(q, \theta) =  e^{-i\theta q}\, \psi_{k+\theta, \lambda}(q, 0) \qquad {\rm and}
\qquad     E_{k, \lambda}(\theta) =  E_{k+\theta, \lambda}(0)\ .
\eea
Thus the   only effect of  $\theta$  in the non-compact case
 is to  reshuffle  the eigenvalues in each energy band  by  
 a universal shift of $k$ -- a trivial effect as stated above.

 The situation  is  different in the compact case where  $P$ is 
a gauge transformation, so it  must  leave
invariant all physical states. We must now project to $k=0$, 
  i.e. keep only the periodic wavefunctions  $\psi_{0, \lambda}(q, \theta)$ whose energy is
$E_{0, \lambda}(\theta)= E_{\theta, \lambda}(0)$. 
 This  keeps one state in each energy band 
 as  illustrated in the figure. If we use as reference the non-compact theory with $\theta =0$,
then gauging the $\mathbb{Z}$  symmetry has the effect of projecting onto 
 the sector with quasi-momentum equal to $\theta$.

More generally, one may   gauge 
 a  subgroup 
  $\{q\to q+  n N\vert n\in \mathbb{Z}\}$ of the global symmetry
 and project  onto   
  $N$ states  with equally-spaced quasi-momenta  in each energy   band.
In the language of the helix, this amounts to taking $z$  periodic  ($z\sim z+1$)
and $\xi = 1/N$ so that the helix has a total of $N$ turns.  
For  $N\gg 1$  the bands are filled very  densely.
{ Alternatively, we  may perturb the infinite-helix model by  a shallow potential 
$\Delta V \sim \epsilon z^2$ with $\epsilon\ll 1$.  Its effect is to   cut  off effectively 
the number of helix turns   to $N\sim 1/\sqrt{\epsilon}$, so the  energy spectrum 
is to a good approximation the same as for a long periodic helix. 
We will see  that both models are realized in gauge theory.}


   One final  comment concerns the width of the bands,  which measures  the 
$\theta$-dependence of the spectrum in the circle model.  For the low-lying  bands 
this is proportional to the amplitude of tunneling between nearby
potential wells  which is  exponentially  small.   At higher energies
the particle gets  delocalized and the bandwidth grows. 
 A  frequently used quantity is the
 topological susceptibility  $\chi$, 
  which is the second derivative of the ground-state energy at $\theta =0$.
This vanishes when tunneling is suppressed,   and it is exponentially small
for a    potential 
 barrier that is   hard to penetrate,
   $\chi \sim \exp(-S_0)$  where   $S_0$ is  the instanton action. For a freely-moving  particle
    $\chi =  1/M$.\,\footnote{The ground state energy of a free particle 
  is   $\theta^2/2M$. It    has a cusp at  
$\theta = \pm \pi$, a fact that can be  attributed more generally to the existence  of many degenerate states
 \cite{Witten:1998uka}.  
We thank Cesar Gomez for this remark. 
 }

 
\section{Chern-Simons Number as Helix Angle}
\label{3}

  The periodic variable in Yang-Mills theory is the Chern-Simons number of the gauge field. 
In temporal  gauge, appropriate for Hamiltonian quantization, this is defined  
 by the well-known expression
 \bea\label{CS}
n_{\rm CS}  =     -{1\over 8\pi^2} 
 \int d^3x\  \epsilon^{ijk}\,   {\rm tr}( A_i \partial_j A_k + {2\over 3} A_i A_jA_k) 
\, :=\,  \int_{{\rm S}^3} {\cal C}(A)
\ ,  
\eea   
where ${\cal C}(A)$ is the shorthand symbol for the normalized Chern-Simons 3-form and we
have  taken 
space to be the 3-sphere  S$^3$. 
The Chern-Simons number is invariant under gauge transformations that are homotopic to 
the identity, 
but it transforms   under large ones by integer shifts, 
$n_{\rm CS} \ \to\  n_{\rm CS}  + n$,  where $n$ is the winding number of the gauge  transformation. 
Since {\it all} of  the gauge transformations must  
 act trivially on physical states,  
   $n_{\rm CS}$ lives on the  circle  and is hence compact. 

  Let us now assume,  following Tye and Wong    \cite{Tye:2015tva}\cite{Tye:2016pxi},   that it
makes sense to write down an effective action for the variable $n_{\rm CS}$. 
We will comment on the validity of this assumption in the end. For now we
only require two topological features of the action: that it is invariant under
integer shifts,   and that its potential be bounded. The maximum of the potential  in the electroweak
theory is the sphaleron's mass  \cite{Klinkhamer:1984di}. 
In the  pure gauge theory considered here the infrared cutoff is  the inverse sphere radius 
that we set equal to one, so the potential barrier  should be of order  $\sim 1/g^2$. 

  For the sake of concreteness, we may write down an action by 
truncating the theory on S$^3$  to the SO(4)-invariant sector.
This reduces configuration space to a single variable, which can be
chosen conveniently to be  $n_{\rm CS}$. The action is then fixed by requiring
that the standard Belavin {\it et al}   instanton \cite{Belavin:1975fg} be a solution of
 the reduced model  in imaginary time \cite{Vandoren:2008xg}.
 The exercise is worked out  in appendix \ref{app}\,. 
The result is more easily  expressed after a non-linear redefinition 
of the Chern-Simons variable, 
\bea\label{3.2}
n_{\rm CS}(q)  = q^2(3-2q)\,,
\eea
where  $q$ must be restricted to the range  $[0, 1]$  and 
\bea\label{3.3}
S(q, \dot q) \
  = \  {12 \pi^2 \over g^2}\int  \hskip -0.6mm dt\,  \left[\,\dot q^{\,2} - 4 q^2 (q- 1)^{\,2}\,\right] -  \theta \int \hskip -0.6mm dt\,  
 \dot n_{\rm CS}(q)  \ . 
\eea
 Note that as $n_{\rm CS}$
  varies from $0$ to $1$, the variable $q$  
covers monotonically this same range.  Since the endpoints of the interval are identified, one  
must glue together the two minima of the 
 double-well potential    keeping the finite  barrier in their middle.  
The resulting potential  is  non-analytic  at the   identification point $0 = q =1$, 
but  it remains   twice-continuously differentiable there. 

 The above action has the same gross  features as that in  \cite{Tye:2015tva}, namely a
harmonic potential around $q=0$ and a barrier height of order 
  $\sim 1/g^2$ at weak coupling.
 Since $q$ is a compact  variable, the energy spectrum  
is  discrete with no bands. Furthermore,  
in contrast to the non-compact toy model  of section \ref{2}\,  here we cannot  decree
 that  the shift symmetry  
is  global rather than gauged. Doing this  violates  {\it locality}, 
 because   gauge transformations in the topologically trivial ($n =0$)
sector can   look `large'  in two widely-separated regions.
 
\medskip

 One  way out 
is to associate  to instanton transitions a shift of  some  bosonic zero mode\,\footnote{Fermionic
zero modes will not do the job, since they at best  double the effective period of $n_{\rm CS}$.} 
thereby turning configuration space into a helix.
Interestingly, this can be achieved without violation of  locality by  coupling the original 
 Yang-Mills   to a topological theory,  as proposed by Seiberg \cite{Seiberg:2010qd}. 
The extra term in  the   Lagrangian  is 
\bea\label{3.4}
\Delta {\cal L}_{\rm YM} \,  =  \,   a\,\left[ d{\cal C}(A) - \xi^{-1} dB\,\right]\ , 
\eea
where $a$ is a Lagrange multiplier field, $B := {1\over 3 !} B_{\mu\nu\rho} dx^\mu\wedge
dx^\nu\wedge dx^\rho$ is a 3-form gauge field, $d$ is the  exterior derivative,  
and $\xi$    is a free parameter -  the pitch of the helix.  
\smallskip

The quantum-mechanical model on S$^3$  has now two new
variables, $a$ and the 
  holonomy of the 3-form field, $b := \int_{S^3}B$. 
Their action  is
\bea\label{delta}
  \Delta S =    \int \hskip -0.6mm dt\  a  \, 
 [\, \dot n_{\rm CS} -   \xi^{-1}\, \dot b\,]\ .  
\eea
The  equation of motion for $b$  forces $a$ to be  constant, 
 while the   equation of  $a$ 
implies\, $\xi \dot n_{\rm CS} =  \dot b$.  As expected, 
this  modification did not add any  {\it dynamical}  degrees of freedom to the theory. 
However, instanton transitions are now accompanied by a shift $\Delta b = \xi \,\Delta n_{\rm CS}$,
which converts  the $n_{\rm CS}$ circle to a helix. 

    We may consider   $b$ to be  a periodic variable, $b\sim b+1$,  if we decree that  the theory
only admits  membranes  whose charge is an integral multiple of $2\pi$. By analogy with  
 conventional gauge theories, this means that the admissible volume operators
(counterparts of Wilson-loop operators) are labeled by an integer charge $m$, 
\bea
  {\cal W}_m :=   e^{2\pi i m \int_{S^3}B}\ . 
\eea
The spectrum of the  model with action $S+\Delta S$, eqs.\,\eqref{3.3} and \eqref{delta}, 
depends then on the  
parameter $\xi$. If $\xi=1$ it is the same as for pure Yang-Mills theory. If $\xi$
is irrational the helix is non-compact and the spectrum has continuous bands.
Finally, if  $\xi  = 1/N$ the helix closes after  $N$   turns and 
 one should keep  $N$ eigenstates
in every  band.

    Let us pick  $\xi = 1/N$.  The energy eigenvalues in the notation
of section \ref{2}\, are  $E_{\theta + 2\pi k/N,\, \lambda}(0)$  where $\lambda$ is the 
label of the band and 
$k=0, 1, \cdots , N-1$. Thus $N$ different $\theta$  sectors coexist and the partition function
of the theory can be written as 
\bea
Z =   \sum_{k=1}^N \,  Z_{\rm YM}\bigl(\theta + {2\pi k\over N}\bigr) \ , 
\eea
where $Z_{\rm YM}(\theta)$ is the partition function of the original Yang-Mills  
at angle $\theta$. This  agrees with the results of Seiberg \cite{Seiberg:2010qd}.
 Note that local operators do not mix the
different  $\theta$ sectors -- these are superselection sectors in the usual sense. 
They are however mixed
by the   operators  ${\cal W}_m$ which shift  $\theta$ 
 by $\Delta\theta = 2\pi m/N$,  as the reader can  verify.

   The   ${\cal W}_m$ are  examples of 
boundary or interface operators (also called Janus interfaces when they carry
no degrees of freedom, as in this case). They  have been studied 
 extensively  in   ${\cal N}=4$ super Yang-Mills  where they preserve space-time
supersymmetry if   combined with an appropriate discontinuity of the coupling  $g$
\cite{D'Hoker:2007xy}\cite{Gaiotto:2008sd}. More generally, these interfaces can be
conformal  but  not {\it topological}  as should be clear from the fact that they modify
  operator dimensions (for a recent discussion of the $\theta$-dependence of
scaling dimensions in ${\cal N}=4$ super Yang-Mills see  \cite{Beem:2013hha}). 
Using radial quantization,  the conformal interface operators can be defined on
3-spheres  in $\mathbb{R}^4$ 
and  expanded for small 3-spheres in  terms of
local operators,   the leading of which is   the topological density  tr$(F\wedge F)$. 
\smallskip

{   After all the dust has settled the theory looks deceivingly simple. It  is a 
superposition of $N$ distinct gauge theories  with different values of the $\theta$-angle, 
 related by   interface operators. There is however  a non-trivial fact: this coexistence was 
 achieved
by a {\it local}  modification of the field theory action.\,\footnote{The same trick can be used to force
the coexistence of 2D sigma models with  a K\"ahler modulus $\tau$ whose real part takes several
different values. In free-field
models the expansion of  interface operators in terms of local operators  
 can be computed  to all orders from the exact conformal boundary states, see e.g. \cite{Bachas:2012bj}.
Interestingly, since the dimension of the identity operator vanishes for all values of $\tau$,
the `bandwidth' is zero for the lowest band but nor for those with non-trivial winding.
} 
The construction is reminiscent of the non-dynamical 3-form introduced by Brown and Teitelboim
to describe a multi-valued   cosmological constant \cite{Brown:1987dd}.
}


\section{Dynamical Axion}
\label{4}

    The 3-form model  of  the previous section  may look  contrived, and one 
can  worry whether  it  arises  from a fundamental theory like string theory.  
It is therefore   interesting  
to see that a similar band structure    emerges  from  a more conventional theory,    
 Yang-Mills coupled   to a dynamical axion field.
The extra  terms in the reduced  action   are
 \bea\label{3.7}
  \Delta S =   \int \hskip -0.6mm dt\,  \left[    \, 
  {f_a^2\over 2}\,  \dot a^2   -   a\,\dot n_{\rm CS}   \right] \  
\eea
where  $f_a$  is the  axion decay constant which we will assume  large.  
We also assume that $a$ is periodic (this is automatic in string-theory embeddings, 
and maybe more generally  \cite{Banks:2010zn})
so the target space  of our quantum mechanical model
is the two-dimensional torus $(a, n_{\rm CS}) \sim (a, n_{\rm CS}+1) \sim (a + 2\pi, n_{\rm CS})$. 
Note that  $2\pi$ is the minimal period  compatible  with a well-defined path-integral measure,
but in general  the period of $a$  can be  $2\pi k$ with $k\in \mathbb{Z}$.\,\footnote{To see why, 
 consider a  gauge  group  $[SU(2)]^k$  broken spontaneously to 
   its diagonal  subgroup. 
The   $ a \dot n_{\rm CS}$\, term in $\Delta S$  is multiplied by  $k$, 
so if  normalized as in  \eqref{3.7} the axion  has  $k$ times the original period.}
This freedom  will not be important to us, so  we set $k=1$. 
\smallskip

The  action $S+\Delta S$,  eqs.\,\eqref{3.3} and \eqref{3.7}, 
describes a non-relativistic particle moving on a 2-dimensional  torus in the  
 background of   a  magnetic field and a non-trivial potential.   
The torus is parametrized by    $(a/2\pi, n_{\rm CS})$ which we will  denote  for short 
 $(x,y)$.   
The background field   $A  = - 2\pi x\, dy$   corresponds to one unit of magnetic flux through the torus. 
Since the quantization of the  system does not depend on the choice of gauge for the magnetic field,
  we may switch to $A  =  2\pi y\, dx$ which amounts to integrating by parts the 
$\int \hskip -0.4mm a \,\dot n_{\rm CS}$\,  term 
 of the action.  The Hamiltonian is the sum of two terms, 
\bea\label{3.8}
{\cal H} =   {1\over 2 f_a^2}\, [\,p_a -  n_{\rm CS}\,]^2 +  {\cal H}_0(\theta)\  
\eea
 where $p_a$ is the momentum conjugate to $a$,  and ${\cal H}_0(\theta)$ is the Hamiltonian  
  of the pure Yang-Mills theory derived from  the action \eqref{3.3}. 

 Let us  analyze the energy spectrum  in   steps. We will first  ignore the periodicity of 
  $n_{\rm CS}$ and  of the related variable $q$,   allowing  them to take values  
on the   real line. Recall that $q$ was  
 a  redefinition of the Chern-Simons number between successive integers, 
 whose  only merit
was to simplify the kinetic  and potential  energy  in our toy-action $S$. 
 Since there is nothing
sacred about $S$,   the 
fine distinction between $q$ and $n_{\rm CS}$   plays  no special role.  
The periodicity of  $n_{\rm CS}$ and  $q$ will be imposed  in the  end. 

Next,  we note that  ${\cal H}$ commutes with the axion momentum $p_a$ which is furthermore quantized,  so we may  set $p_a = n_a \in\mathbb{Z}$.  The problem now 
reduces  to that of a particle (with canonical kinetic energy and mass $\sim 1/g^2$)  moving 
in   the  effective   1D potential
\bea\label{V} 
  V(q, n_a) \ = \   {1\over 2 f_a^2}\, \left[\,n_a  -  n_{\rm CS}(q)\right]^2 + V_0(q) \ .  
\eea
Here $V_0$ is the potential of the pure Yang-Mills theory which has been 
periodically extended from the interval 
$[0,1]$ to the entire real line.

 We have assumed  that  the axion scale  is much larger than the other scales of the problem, 
 the inverse sphere radius  ($f_a\gg 1$)   and the sphaleron's mass  ($f_a \gg 1/g^2$). 
    The spectrum is then,  to a good approximation, 
 determined by the periodic   potential $V_0$
 which leads to  the characteristic band structure of a lattice model. 
The only effect  of the axion  term in  
 \eqref{V}  is to put the particle  inside a large box, centered around the Chern-Simons  number
$ n_{\rm CS}  = n_a$ and 
of approximate size 
$f_a $. The   box discretizes  the quasi-momenta, but since  their spacing
$\sim 1/f_a$ is  small   they  cover  the energy bands rather densely.  
\smallskip

 We can finally restore the winding transformations to their  status of {\it gauge}, 
 not {\it global}  symmetries. In   pure Yang-Mills  this
  identified  periodically the Chern-Simons number $n_{\rm CS}$.
But   \eqref{V}  is only invariant
under the  combined shifts  $(n_{\rm CS}, n_a) \to (n_{\rm CS}+n, n_a+n)$, so this  
must be  the symmetry that we need  to gauge. This  is familiar from the
Landau problem on the  square torus $(x,y) \sim (x+1, y)\sim (x, y+1)$. 
The free-particle Hamiltonian in the $A_x = By$\, gauge is $
 {\cal H} =  {1\over 2} (p_x - By)^2 + {1\over 2} p_y^2$. It  commutes with the  
torus translations
  \bea
U= e^{ip_x}\qquad {\rm and}\qquad V = e^{ip_y} e^{-i Bx} \ ,  
\eea
which furthermore commute with each other if  $B$
 obeys the Dirac quantization  $B/2\pi =k  \in  \mathbb{Z}$. 
The extra factor in   $V$ is the gauge transformation 
that  patches  together the local charts at  $y$ and   $y+1$.  
The torus identifications impose on  wavefunctions the conditions  $U\psi = V\psi =\psi$, and 
these  are indeed obeyed by  the $k$ independent states  at each Landau level, 
\bea
\psi_{\lambda, n_x}(x,y) =  \sum_{n= -\infty}^\infty   e^{2\pi i(n_x+ k n) x}\,  \psi_\lambda \bigl(y - n - {n_x\over k}\bigr) \ , 
\eea
 where  $\psi_\lambda(y)$ are the harmonic-oscillator wavefunctions at level $\lambda$, 
and $p_x/2\pi  =  n_x \in  \mathbb{Z}\,$(mod\,$k$). 

\smallskip 

     Coming back to our problem we conclude  that the large gauge transformations identify
periodically $(n_{\rm CS},  n_a) \sim (n_{\rm CS}+1,  n_a+1)$, thus converting again the
configuration space  into a helix. 
The second helix coordinate  is the axion momentum $n_a$, which 
   is  only an approximate zero mode   lifted by the small quadratic term in the
potential \eqref{V}. In the limit $f_a\to \infty$ the zero mode becomes exact
and the  splitting  of the quasi-momenta vanishes.

    That the axion theory  gives a similar  spectrum  as the topological model of the previous 
section is,  actually,  no  coincidence. We could have argued for this directly by   adding   to the Lagrangian \eqref{3.4} a small
mass term for the 3-form field,  
  ${\cal L}_{\rm mass} =  (\xi f_a)^{-2} B_{\mu\nu\rho}B^{\mu\nu\rho}/12$.
Integrating out  the auxiliary field $B$ gives, after  integration by parts, 
 $B_{\mu\nu\rho} = \xi f_a^2 \, \epsilon_{\mu\nu\rho\sigma}\partial^\sigma a$  thus
  reproducing precisely the dynamical axion model. Since the mass term breaks the
periodicity of $B$,  interface operators of any  charge are now allowed
\bea
 {\cal W}_\gamma =  e^{i\gamma \int_{S^3} \,*da}\qquad \forall \gamma\in\mathbb{R}\ .  
\eea
It is worth stressing that an approximate band structure would also arise from
an axion potential with a large number of 
  stable minima between $0$ and $2\pi$, but such a model
would be really  contrived. 
The simple axion model considered here is the one emerging naturally from string theory, 
with $f_a$  
of order the Planck scale .
 \smallskip

  {  We should here pause to assess the validity of our approximations. Reducing
  the gauge theory on S$^3$ to a quantum-mechanical model for $n_{\rm CS}(t)$ has, a priori,
zero range of validity.  Indeed, the perturbative excitations of this latter model have energy $\sim 1$ (the
inverse sphere radius), which is also  
 the energy of the truncated Kaluza-Klein modes. The only thing that this model  accurately  describes
is the $\theta$-dependence of the vacuum energy, or  more precisely
 its  exponential sensitivity $e^{-S_0}$. This is, however,  sufficient  for our purpose  here,
which was to explain  
how band structure may emerge in Yang-Mills. 
Note that  level-splittings  in the lowest band are of order $ f_a^{-1}  e^{-S_0}\sim  N^{-1}  e^{-S_0}$, 
  much smaller than the  infrared cutoff which is also the gap between bands. 
}


\section{Concluding Remarks}
\label{5}

  The present  investigation  was prompted by ref.\,\cite{Tye:2015tva} which challenged the earlier
consensus   that   $B+L$-violating  electroweak processes should be invisible
at   LHC.\,\footnote{A
characteristic signature  
would be the   large number  of electroweak  bosons produced in 
the decay of the sphaleron  \cite{Gibbs:1994cw}\cite{Ellis:2016ast}\cite{Ellis:2016dgb}. This is similar to the decay of
hypothetical  black holes in low-scale quantum gravity models 
 for which the ATLAS collaboration has   published  exclusion plots in
\cite{Aad:2015mzg}.}\,   
The standard wisdom  is that the collision energy, initially carried by two hard quanta, has no
chance of being streamlined to a coherent excitation of one (or a few)  degrees of freedom
in a weakly-coupled theory like the electroweak Standard Model. 
Thus all processes involving solitons as intermediate states
should be exponentially suppressed,  independently of any other details of the theory at hand.

This `wisdom' is supported by a lot of theoretical evidence, and  also experimental 
\cite{Ellis:2016ast}\cite{Ellis:2016dgb}. In the context of the simple quantum mechanics of the
Chern-Simons number analyzed here, the suppression would not come from the tunneling
rate of an excited state, but rather from the unlikely probability of reaching   this state when  the energy
is  initially stored  in a `hard quantum' that  can be modelled by a highly-energetic source linear in $q$ \cite{Bachas:1991cb}\cite{Bachas:1992dw}.
  Still, in the absence of a definitive calculation and since the LHC Run2 
 operates  in the sphaleron  range, it is important to scrutinize all `heretic' proposals. 

\smallskip

   We focussed here  on one of the issues raised in ref.\,\cite{Tye:2015tva}, 
namely   whether the electroweak  theory exhibits  band structure similar to  that of an electron in a solid. 
We have explained why this  is not  the case, although it would be 
 conceivable   if 
Yang-Mills  theory were coupled to an  axion field with high Peccei-Quinn scale and 
 if space could be treated as being compact. 
 It is amusing to note that even this hypothetical  band structure  would be
absent  in the electroweak theory for a different  reason:  Because  
 $\theta$  can be rotated away by a chiral $U(1)_B$ and/or $U(1)_L$ rotation.
In contrast to what happens for QCD  such chiral rotations do not 
transfer the electroweak $\theta$ angle  
to  the  Yukawa couplings of  quarks and leptons, so $\theta$  is  a relevant parameter
only if  both $B$ and $L$ are explicitly broken
 \cite{Anselm:1993uj}.

\vskip 1cm
\noindent {\bf Aknowledgements} 
\smallskip

\noindent   We thank  Ilka Brunner, Cesar Gomez, Slava Mukhanov, Giuseppe Policastro 
 and Jan Troost for discussions. T.T. thanks the Philippe Meyer Institute of the 
\'Ecole Normale  for support during a
visit that initiated this project. C.B.  thanks the  LMU physics department for hospitality during
completion of this work. The work of T.T. is partially supported by the European Union Seventh Framework Program (FP7-REGPOT-2012-2013-1) under grant agreement No. 316165. 
 
\vfil\eject



\appendix
\section{An Action for the Chern-Simons number}
\label{app}

A  prototypical action for the Chern-Simons number $n_{\rm CS}$ can be found by 
putting  Yang-Mills theory on   S$^3\times\mathbb{R}_\tau$ where 
S$^3$ is the unit-radius  round 3-sphere  and $\tau$ is imaginary time. 
This  spacetime is related to  Euclidean $\mathbb{R}^4$
by a Weyl transformation,
\bea
    ds^2[{\rm S}^3\times\mathbb{R}]  = d\tau^2 +   d\Omega^2_{3} \ = \ {1\over r^2} ( dr^2 + r^2 d\Omega^2_{3}) \  
=   \ {1\over r^2}\ ds^2[\mathbb{R}^4]  
\eea
with  $\tau =   \log r$,   and  $d\Omega^2_{3}$ the metric on the round 3-sphere. 
Since the self-dual Yang-Mills equations are Weyl-invariant,\,\footnote{The self-duality
 equations are  $F_{\mu\nu} = {1\over 2}
\epsilon_{\mu\nu}^{\ \ \ \rho\sigma} F_{\rho\sigma}$,  where the Levi-Civita tensor obeys
 ${  \sqrt{\vert g\vert}} \, \epsilon^{\mu\nu\rho\sigma} \equiv 
\hat\epsilon^{\mu\nu\rho\sigma}$ with  $\hat\epsilon$  the totally antisymmetric symbol normalized to $\pm1$.
Clearly $\epsilon_{\mu\nu}^{\ \ \ \rho\sigma}$ is Weyl invariant. } 
 all solutions  in 
$\mathbb{R}^4$  also solve the equations in S$^3\times \mathbb{R}$. 
This is  in particular true  for the celebrated  Belavin {\it et al}  \cite{Belavin:1975fg}  instanton 
\bea
 A_\mu   =  -     { \bar \sigma_{\mu\nu}  (x^\nu-a^\nu) \over (x-a)^2 + \rho^2}\  \quad \Longrightarrow 
\quad   F_{\mu\nu}    = { 2 \bar \sigma_{\mu\nu} \,   \rho^2  \over [(x-a)^2 + \rho^2]^{\,2} }\    
\eea
where $ \bar \sigma_{\mu\nu} = -i \eta^a_{\mu\nu} \sigma^a$\,,   $\eta$ is the 't Hooft tensor
and $\sigma^a$ the Pauli matrices.  
We are following here the conventions of reference \cite{Vandoren:2008xg}. 
The   instanton center,    $a^\nu$,  and  the instanton
scale, $\rho$,  are collective coordinates of the solution. 
 
    To express this solution in terms of $\tau$ and coordinates   $\phi_m$ on the 3-sphere    
we use the relation  $x^\mu = \vert x\vert\, \hat n^\mu =  e^{\tau} \hat n^\mu(\phi_m)$ where 
$\hat n_ \mu \hat n^\mu = 1$. The instanton centered at  $a^\nu =0$, which preserves an O(4) symmetry,   
 is automatically  in the $A_\tau = 0$ gauge thanks to the antisymmetry of $ \bar \sigma_{\mu\nu}$.
The vector potential, and  the electric and magnetic fields  read 
\bea\label{A3}
  A_m =    { \bar \sigma_{\mu\nu}\,  \hat n^\mu  {\partial_m \hat n^\nu }
\over  1  +  e^{2(\tau_0  -\tau )}}\  , 
 \quad   F_{\tau m}    = { \bar \sigma_{\mu\nu}\,  \hat n^\mu  {\partial_m \hat n^\nu }
\over  2 \cosh^2(\tau - \tau_0)    }\  , \quad  
 F_{l  k }    = {  \bar \sigma_{\mu\nu}\, \partial_l  \hat n^\mu  {\partial_k \hat n^\nu }
\over  2 \cosh^2(\tau - \tau_0)    }\ .
\eea
We have here defined $\rho \equiv e^{\tau_0}$ to make clear that the instanton scale in $\mathbb{R}^4$
becomes the 
time collective coordinate in the new coordinate system.\,\footnote{This is familar from holographic dualities,
where one identifies the instanton moduli with the coordinates of a D-instanton in AdS$_5$. 
The moduli  $a^\nu$ and $\rho$ are Poincar\'e coordinates 
for  AdS$_5$, while $\tau_0$ is the time in global coordinates. Setting $a^\nu=0$ amounts to
putting the D-instanton in the center of global AdS$_5$. 
} 
The gauge field  vanishes at $\tau \to -\infty$ and is pure gauge at  $\tau \to +\infty$, so its 
  field strength vanishes at both ends as expected since it 
 describes vacuum-to-vacuum tunneling.

From now on we set $\tau_0 =0$ and think of \eqref{A3} as a path in configuration space parametrized by $\tau$. 
 The potential energy along the path is the energy stored in the magnetic fields,
\bea
V(\tau) =  -{1\over 2g^2} \int_{{\rm S}^3} {\rm tr} (F_{lk} F^{lk} ) =  {3\pi^2 \over g^2}\, {1\over \cosh^4\tau}\ , \qquad 
 V_{\rm max} =  {3\pi^2 \over g^2}\ .  
\eea
The normalization can be quickly  fixed by noting that the  integral of $V$  is half the instanton action, 
$\int d\tau V(\tau) = {4\pi^2\over g^2}$. The energy barrier is highest at $\tau =0$,  the analog
of the sphaleron of the Standard Model. 
 Note that the role of the W-boson mass   is   played by the radius of
the 3-sphere,  which introduces a cutoff on the instanton size. In holographic language  the barrier height
attains its (non-zero)  minimum when the D-instanton sits  at the center of AdS$_5$. 
\vskip 2mm
\smallskip

 One may generalize the gauge field \eqref{A3} to a variational ansatz (in temporal gauge)  that
depends on a single  arbitrary function $y(\tau)$,
\bea
 A_\tau = 0\ , \qquad A_m =    { \bar \sigma_{\mu\nu}\,  \hat n^\mu  {\partial_m \hat n^\nu }
\over  1  +  e^{ -2y(\tau)}}\ . 
\eea
  It can be checked  that this is the most general SO(4)-invariant configuration of the gauge fields,
so this  is a symmetry truncation of configuration space. There is a single surviving
degree of freedom, which we can take to be the Chern-Simons number $n_{\rm CS}$.
  The magnetic fields for the above  configuration are the same as in \eqref{A3} 
with $\tau$ replaced by $y(\tau)$, while
the electric fields are also multiplied by $\dot y$. 
 Notice that this is {\it not} a   reparametrization since we keep the same metric $g_{\tau\tau} = 1$.
 Computing the energy of the electric and magnetic fields gives
the effective   Lagrangian (in Euclidean time)\,\footnote{The general Lagrangian that admits $\dot y=1$
as a solution is ${\cal L} = f(y) (\dot y^2 +1)$. Our calculation of the energy stored in the magnetic fields  fixes the arbitrary function $f(y)$.} 
\bea\label{A6}
{\cal L}(q, \dot q) =  {3\pi^2 \over g^2}\, {1\over \cosh^4 y} (\dot y^2 + 1)\ . 
\eea
One can check  that  $\dot y = 1$ solves  the variational equation for this action, so the instanton
is a solution as it should be. The change of variable $z = \tanh y$ transforms the problem  to
the standard double-well quantum mechanics, 
\bea\label{A7}
{\cal L}(q, \dot q) =  {3\pi^2 \over g^2}\, \left[ \dot z^2 + (z^2-1)^2\right]\ . 
\eea

 Finally we can relate the variable $z$ to  the Chern-Simons number of the field, 
\bea\label{A7}
 n_{\rm CS}(q)  =  - {1\over 8\pi^2}
\int_{{\rm S}^3}  \epsilon^{lkm}\, {\rm tr} ( A_l \partial_k A_m + {2\over 3} A_l A_k A_m)\ . 
\eea
Using the fact that space is compact and that
$n_{\rm CS}(-\infty) =0$, one can rewrite the above integral  as a 4-dimensional  integral  
of  
${\rm tr} (F_{\mu\nu}\,^*F^{\mu\nu})$. This  does not involve the metric, so
the result is form-invariant  whether  expressed as function of  $\tau$ or $y$. 
Performing the integral  gives
 \bea
n_{\rm CS}  =  {1\over 2} + {\sinh(3y) + 3 \sinh y \over 8 \cosh^3y}\ =
\ {1\over 2} + {z\over 4} (3-z^2)
\ . 
\eea
In principle, one can invert this relation  to  find an effective quantum mechanics  
 \bea\label{A10}
{\cal L} = {1\over 2} M(n_{\rm CS})\,  \dot n_{\rm CS}^2 - V(n_{\rm CS})
\eea
with functions $M$ and $V$ that can be computed. 

Note that the range of $z$  is $[-1, 1]$, and the corresponding range of $n_{\rm CS}$
is $[0, 1]$, so the
  Lagrangians \eqref{A7}  and \eqref{A10} 
should be restricted to these finite  intervals. A last change of variable to $q = (z+1)/2$,
and Wick rotation to real time, 
gives finally the   expressions \eqref{3.2} and \eqref{3.3} of section 3.
   The range of  $q$ is the  same  as the range of $n_{\rm CS}$, so $q$  can be considered as
a convenient redefinition of the Chern-Simons number. Since the only features 
of the action  that are relevant to  our discussion are (1) its periodicity, 
and (2) the fact that  the  potential barrier traversed by winding histories is finite, the fine
distinction between $n_{\rm CS}$ and $q$ is not  important.  
 \vskip 2mm

 \vfil\eject


%

\end{document}